\documentclass[12pt,twoside]{article}
\usepackage{psfig}
\newcommand{\VolumeHeader}{}
\newcommand{\VolumeSerial}{LNS}

\newcommand{\ActivityName}{ {\normalsize {\it 
Gravitational Waves: A Challenge to Theoretical Astrophysics
}}}
\newcommand{\ActivityDate}{ {\normalsize {\it
Trieste, 5-9 June 2000 
}}}

\newcommand{\be}{\begin{equation}}
\newcommand{\ee}{\end{equation}}
\newcommand{\bea}{\begin{eqnarray}}
\newcommand{\eea}{\end{eqnarray}}
\newcommand{\LectureHeader}{Deviations from...}

\begin{document}
\pagestyle{myheadings}
\markboth{\LectureHeader}{\VolumeHeader}
\markright{\VolumeHeader}

%%%%%%%%%%%%%%%%%%%%%%%%%%%%%%%%%%%%%%%%%%%%%%%%%%%%%%%%%%%%%%%%%%%%%%%%%%%
%%%            Title page starts here                                     %
%%%%%%%%%%%%%%%%%%%%%%%%%%%%%%%%%%%%%%%%%%%%%%%%%%%%%%%%%%%%%%%%%%%%%%%%%%%

\begin{titlepage}

%%% YOUR CHANGES BELOW THIS LINE

\title{Deviations from Einstein's Gravity at Large and Short Distances} 

\author{Francesco Fucito\thanks{fucito@roma2.infn.it}
\\[1cm]
{\normalsize
{\it I.N.F.N. sez. Roma 2, Via della Ricerca Scientifica, Rome, Italy.}}
\\[10cm]
%%% FOR FURTHER AUTHORS SEE WHAT IT IS WRITTEN IN THE ABSTRACT 
%%% DO NOT CHANGE THE FOLLOWING LINES
{\normalsize {\it Lecture given at the: }}
\\
\ActivityName 
\\
\ActivityDate 
\\[1cm]
{\small \VolumeSerial} 
}
\date{}
\maketitle
\thispagestyle{empty}
\end{titlepage}

\baselineskip=14pt
\newpage
\thispagestyle{empty}

%%%%%%%%%%%%%%%%%%%%%%%%%%%%%%%%%%%%%%%%%%%%%%%%%%%%%%%%%%%%%%%%%%%%%%%%%%%
%%%            Abstract page starts here                                  %
%%%%%%%%%%%%%%%%%%%%%%%%%%%%%%%%%%%%%%%%%%%%%%%%%%%%%%%%%%%%%%%%%%%%%%%%%%%

\begin{abstract}

%%% YOUR CHANGES BELOW THIS LINE
In this talk i will describe some recent results on the sensitivity
of resonant mass detectors shaped as a hollow sphere to scalar gravitational
radiation. Detection of this type of gravitational radiation will signal
deviations from Einstein's gravity at large distances. I will then discuss
a class of experiments aiming at finding deviations from Einstein's gravity
at distances below 1 cm. I will review the main experimental difficulties
in performing such experiments and evaluate the effects to be taken in account
besides gravity.
\end{abstract}

\vspace{6cm}



%%%%%%%%%%%%%%%%%%%%%%%%%%%%%%%%%%%%%%%%%%%%%%%%%%%%%%%%%%%%%%%%%%%%%%%%%%%
%%%       Automatic TOC and your Text starts here                         %
%%%%%%%%%%%%%%%%%%%%%%%%%%%%%%%%%%%%%%%%%%%%%%%%%%%%%%%%%%%%%%%%%%%%%%%%%%%

\newpage
\thispagestyle{empty}
\tableofcontents

\newpage
\setcounter{page}{1}

\section{Introduction}
It seems reasonable to predict that the new gravitational wave (GW) detectors
now under construction, once operating at the maximum of their sensitivity,
will be able to detect GWs. Will it be possible to use these future
measurements to try to gain information on which is the theory of gravity
at low energies? There are no particular reasons, in fact, why GW must be
of spin two. In reality, many theories of gravity can be built which contain
scalars and vectors. These theories are mathematically well founded.
String theory, more in particular, is believed to be consistent also as
a quantum description of gravity. The predictions of these theories must
then be checked against available experimental data. This forces the
couplings and masses present in the Lagrangian to take values in well
defined domains. See \cite{will} for a more detailed exposition. Once
detected, one can also attempt to use GWs as a mean to further constrain
this picture. It seems to us relevant to try to develop the theory to the
point where it can profit from new experimental insights. For these reasons
the interaction and cross section of a massive elastic
sphere with scalar waves has been analysed in great detail in 
reference \cite{bian-98} ---see also\cite{lobo-95,wp-75}. 

An appealing variant of the massive sphere is a {\it hollow\/} sphere
\cite{vega-98}. The latter has the remarkable property that it enables the
detector to monitor GW signals in a significantly {\it lower frequency range\/}
---down to about 200 Hz--- than its massive counterpart for comparable
sphere masses. This can be considered a positive advantage for a future
world wide network of GW detectors, as the sensitivity range of such
antenna overlaps with that of the large scale interferometers, now in
a rather advanced state of construction \cite{ligo,virgo}. 
A  first study of the response of such a detector 
to the GW energy emitted by a binary system constituted of stars of masses 
of the order of the solar mass was performed in \cite{brun-99}. 
A simple formula for the GW energy was 
obtained in the Newtonian approximation whose region of validity
encompasses emitted frequencies of the order of the first 
resonant mode of the detector under study.

A hollow sphere obviously has the same symmetry of the massive one, so
the general structure of its {it normal modes\/} of vibration is very
similar in both \cite{vega-98}. In particular, the hollow sphere is
very well adapted to sense and monitor the presence of scalar modes in
the incoming GW signal. 

In the first part of this talk I will report on the results of an extension
of the analysis of the response of a hollow sphere, to include scalar
excitations\cite{cfls}. 

The second part of this talk will be dedicated to a class of recently proposed
experiments to measure deviations from Einstein's gravity at distances
smaller than 1 cm \cite{lcp}. The experimental techniques involved in 
performing such experiments are tightly connected to those employed in the
detection of GW from resonant mass detectors. I will review the main
experimental difficulties in trying to achieve a useful signal-to-noise ratio,
estimating the various backgrounds which could screen the gravitational signal.

\section{The hollow sphere}
\subsection{Review of hollow sphere normal modes}

This section contains some review material which is included essentially to
fix the notation and to ease the reading of the ensuing sections. Notation
will be that of reference \cite{vega-98}. The eigenmode equation for a
three-dimensional elastic solid is the following:

\begin{equation}
\nabla^2{\bf s}+\left(1+\lambda/\mu\right)
\nabla (\nabla\cdot{\bf s})=-k^2{\bf s}\ ,\qquad
\left(k^2\equiv\varrho\omega^2/\mu\right),
\label{1.1}
\end{equation}
as described in standard textbooks, such as \cite{love-44,landau-70}. The
equation must be solved subject to the {\it boundary conditions} that the
solid is to be free from tensions and/or tractions. In the case of a hollow
sphere, I have two boundaries given by
the outer and the inner surfaces
of the solid itself. I use the notation $a\/$ for the inner radius, and $R\/$ for
the outer radius. The boundary conditions are thus expressed by

\begin{equation}
\sigma_{ij}n_j=0\hspace{1cm}\mbox{at}\hspace{0.4cm} r=R
\hspace{0.4cm}\mbox{and at}\hspace{0.4cm}r=a\hspace{0.5cm}(R\geq a \geq 0),
\label{bc}
\end{equation}
where $\sigma_{ij}\/$ is the stress tensor, and is given by \cite{landau-70}

\begin{equation}
\sigma_{ij}=\lambda\,u_{k,k}\, \delta_{ij}\,+\,2\,\mu \,u_{(i,j)}.            
\end{equation}
with $\lambda$ and $\mu$ the material's Lam\'e coefficients, and {\bf n}
the unit, outward pointing normal vector.

The general solution to equation (\ref{1.1}) is a linear superposition of
a longitudinal vector field and two transverse vector fields, i.e.,

\begin{equation}
{\bf s}(r,\vartheta,\phi) =
\frac{C_{\rm l}}{q}\,{\bf s}_{\rm l} + \frac{C_{\rm t}}{k}\,{\bf s}_{\rm t}
+ C_{\rm t'}\,{\bf s}_{\rm t'}
\label{1.25}
\end{equation}
where $C_{\rm l}$, $C_{\rm t}$ and $C_{\rm t'}$ are constant coefficients,
and

\label{1.3}
\begin{eqnarray}
{\bf s}_{\rm l}\left(r,\vartheta,\phi \right) &=& \frac{dh_l(qr,E)}{dr}\,
Y_{lm}{\bf n} - \frac{h_l(qr,E)}{r}\,i{\bf n}\times {\bf L}Y_{lm}
\label{1.3a} \\
{\bf s}_{\rm t}\left(r,\vartheta,\phi\right) &=& -l\left(l+1\right)
\frac{h_l\left(kr,F\right)}{r}\,Y_{lm}{\bf n} + \left[
\frac{h_l(kr,F)}{r} + \frac{dh_l\left(kr,F\right)}{dr}\right)]\,
i{\bf n}\times {\bf L}Y_{lm} \label{1.3b} \\
{\bf s}_{\rm t'}\left(r,\vartheta,\phi\right) &=&
h_l(kr,F)\,i{\bf L}Y_{lm} \label{1.3c}
\end{eqnarray}
with $E\/$ and $F\/$ also arbitrary constants,

\begin{equation}
q^2\equiv k^2\,\frac{\mu}{\lambda+\mu} =
\frac{\varrho_0\omega^2}{\lambda+\mu}
\label{1.35}
\end{equation}
and

\begin{equation}
h_l\left(z,A\right)\equiv j_l(z) + A\,y_l(z)
\label{1.4}
\end{equation}
$j_l\/$ $y_l\/$ are spherical Bessel functions \cite{abram-72}:

\begin{eqnarray}\label{1.5}
j_l(z) &=& z^l\,\left(-\frac 1z\,\frac d{dz}\right)^l\,\frac{\sin z}z
\label{1.5a} \\
y_l(z) &=& -z^l\,\left(-\frac 1z\,\frac d{dz}\right)^l\,\frac{\cos z}z
\label{1.5b}
\end{eqnarray}

Finally, {\bf L\/} is the {\it angular momentum\/} operator

\begin{equation}
{\bf L}\equiv -i\,{\bf x}\times\nabla
\label{1.6}
\end{equation}

The boundary conditions (\ref{bc}) must now be imposed on the generic solution
to equations (\ref{1.1}). After some rather heavy algebra it is finally found
that there are two families of eigenmodes, the {\it toroidal\/} (purely
rotational) and the {\it spheroidal\/}. Only the latter couple to GWs
\cite{bian-96}, so I shall be interested exclusively in them. The form
of the associated wavefunctions is

\begin{equation}
{\bf s}_{nlm}^S(r,\vartheta ,\phi) =
A_{nl}(r)\,Y_{lm}(\vartheta ,\phi)\,{\bf n} -
B_{nl}(r)\,i{\bf n}\times{\bf L}Y_{lm}(\vartheta ,\phi)
\label{1.7}
\end{equation}
where the radial functions $A_{nl}(r)$ and $B_{nl}(r)$ have rather
complicated expressions:

\begin{eqnarray}\label{1.8}
A_{nl}(r) &=& C_{nl}\,\left[\frac{1}{q_{nl}^S}\,\frac{d}{dr}\,
j_l(q_{nl}^Sr) -
l(l+1)\,K_{nl}\,\frac{j_l(k_{nl}^Sr)}{k_{nl}^Sr}+\right.
  \nonumber \\
&& \ \qquad\quad + \left. D_{nl}\,\frac{1}{q_{nl}^S}\,\frac{d}{dr}\,
y_l(q_{nl}^Sr)
- l(l+1)\,\tilde D_{nl}\,\frac{y_l(k_{nl}^Sr)}{k_{nl}^Sr}\right]
\label{1.8a}  \\
B_{nl}(r) &=& C_{nl}\,\left[\frac{j_l(q_{nl}^Sr)}{q_{nl}^Sr} -
K_{nl}\,\frac 1{k_{nl}^Sr}\,\frac d{dr}\left\{r\,j_l(k_{nl}^Sr)\right\} +
\right.  \nonumber \\
&& \ \qquad\quad + \left. D_{nl}\,\frac{y_l(q_{nl}^Sr)}{q_{nl}^Sr} -
\tilde D_{nl}\,\frac 1{k_{nl}^Sr}\,\frac d{dr}\left\{r\,y_l(k_{nl}^Sr)\right\}
\right] \label{1.8b}
\end{eqnarray}

Here $k_{nl}^SR$ and $q_{nl}^SR$ are dimensionless {\it eigenvalues\/},
and they are the solution to a rather complicated algebraic equation for
the frequencies $\omega\/$\,=\,$\omega_{nl}\/$ in (\ref{1.1}) ---see
\cite{vega-98} for details. In (\ref{1.8a}) and (\ref{1.8b}) I have set

\begin{equation}
K_{nl}\equiv\frac{C_{\rm t}q_{nl}^S}{C_{\rm l}k_{nl}^S}\ ,\qquad
D_{nl}\equiv\frac{q_{nl}^S}{k_{nl}^S}\,E\ ,\qquad
\tilde D_{nl}\equiv\frac{C_{\rm t}Fq_{nl}^S}{C_{\rm l}k_{nl}^S}
\label{1.9}
\end{equation}
and introduced the normalisation constant $C_{nl}$, which is fixed by the
orthogonality properties

\begin{equation}
\int_V({\bf s}_{n^{\prime}l^{\prime}m^{\prime}}^S)^*\cdot
({\bf s}_{nlm}^S)\,\varrho_0\,d^3 x =
M\,\delta_{nn^{\prime}}\delta_{ll^{\prime}}\delta_{mm^{\prime}} 
\label{1.10}
\end{equation}
where $M\/$ is the mass of the hollow sphere:

\begin{equation}
M = \frac{4\pi}3\,\varrho_0 R^3\,(1-\varsigma ^3)\ ,\qquad
\varsigma\equiv\frac{a}{R}\leq 1
\label{1.11}
\end{equation}

Equation (\ref{1.10}) fixes the value of $C_{nl}$ through the radial integral

\begin{equation}
\int_{\varsigma R}^R\,\left[A_{nl}^2(r) + l(l+1)\,B_{nl}^2(r)\right]\,
r^2dr = \frac{4\pi}3\varrho_0\,(1-\varsigma^3)R^3.
\label{1.12}
\end{equation}

\subsection{Absorption cross sections}

As seen in reference \cite{lobo-95}, a scalar--tensor theory of GWs such as
JBD predicts the excitation of the sphere's monopole modes {\it as well as
the\/} $m\/$\,=\,0 quadrupole modes. In order to calculate the energy absorbed
by the detector according to that theory it is necessary to calculate the
energy deposited by the wave in those modes, and this in turn requires that
I solve the elasticity equation with the GW driving term included in its
right hand side. The result of such calculation was presented in full
generality in reference \cite{lobo-95}, and is directly applicable here
because the structure of the oscillation eigenmodes of a hollow sphere is
equal to that of the massive sphere ---only the explicit form of the
wavefunctions needs to be changed. I thus have

\begin{equation}
E_{\rm osc}(\omega_{nl}) = \frac{1}{2}\,Mb^2_{nl}
  \,\sum_{m=-l}^l\,|G^{(lm)}(\omega_{nl})|^2
\label{2.0}
\end{equation}
where $G^{(lm)}(\omega_{nl})$ is the Fourier amplitude of the corresponding
incoming GW mode, and

\begin{eqnarray}\label{2.1}
b_{n0} &=& -\frac{\varrho_0}{M}\,\int_a^R\,A_{n0}(r)\,r^3 dr
\label{2.1a} \\[1 ex]
b_{n2} &=& -\frac{\varrho_0}{M}\,\int_a^R\,\left[A_{n2}(r)
	 + 3B_{n2}(r)\right]\,r^3 dr
\label{2.1b}
\end{eqnarray}
for monopole and quadrupole modes, respectively, and $A_{nl}(r)$ and
$B_{nl}(r)$ are given by (\ref{1.8}). Explicit calculation yields

\begin{eqnarray}\label{2.2}
\frac{b_{n0}}{R} &=& \frac 3{4\pi}\,\frac{C_{n0}}{1-\varsigma^3}\,
\left[\Lambda(R) - \varsigma^3\Lambda(a)\right] \label{2.2a} \\[1 ex]
\frac{b_{n2}}{R} &=& \frac 3{4\pi}\,\frac{C_{n2}}{1-\varsigma^3}\,
\left[\Sigma(R) - \varsigma^3\Sigma(a)\right] \label{2.2b}
\end{eqnarray}
with

\begin{eqnarray}\label{2.3}
\Lambda(z) & \equiv & \frac{j_2(q_{n0}z)}{q_{n0}R} +
 D_{n0}\,\frac{y_2(q_{n0}z)}{q_{n0}R}   \label{2.3a} \\[1 em]
\Sigma(z) & \equiv & \frac{j_2(q_{n2}z)}{q_{n2}R} -
3K_{n2}\,\frac{j_2(k_{n2}z)}{k_{n2}R} +
D_{n2}\,\frac{y_2(q_{n2}z)}{q_{n2}R} -
3\tilde D_{n2}\,\frac{y_2(k_{n2}z)}{k_{n2}R}   \label{2.3b}
\end{eqnarray}

The absorption {\it cross section\/}, defined as the ratio of the absorbed
energy to the incoming flux, can be calculated thanks to an {\it optical
theorem\/}, as proved e.g.\ by Weinberg \cite{wein-72}. According to that
theorem, the absorption cross section for a signal of frequency $\omega\/$
close to $\omega_N\/$, say, the frequency of the detector mode excited by
the incoming GW, is given by the expression

\begin{equation}
\sigma(\omega) = \frac{10\,\pi\eta c^2}{\omega^2}\,
		 \frac{\Gamma^2/4}{(\omega -\omega_N)^2 + \Gamma^2/4}
\label{2.4}
\end{equation}
where $\Gamma$ is the {\it linewitdh\/} of the mode ---which can be
arbitrarily small, as assumed in the previous section---, and $\eta\/$
is the dimensionless ratio

\begin{equation}
  \eta = \frac{\Gamma_{\rm grav}}{\Gamma} =
	 \frac{1}{\Gamma}\,\frac{P_{GW}}{E_{\rm osc}}
\label{2.5}
\end{equation}
where $P_{GW}$ is the energy {\it re-emitted\/} by the detector in the form
of GWs as a consequence of its being set to oscillate by the incoming signal. In the following I will only consider the case 
$P_{GW}=P_{\rm scalar-tensor}$ with \cite{lobo-95,bian-98}
\begin{equation}
P_{\rm scalar-tensor} = \frac{2G\,\omega ^6}{5c^5\,(2\Omega _{BD}+3)}\,
\left[\left|Q_{kk}(\omega)\right|^2 +
\frac 13\,Q_{ij}^*(\omega)Q_{ij}(\omega)\right]
\label{2.8}
\end{equation}
where $Q_{ij}(\omega)$ is the quadrupole moment of the hollow sphere:

\begin{equation}
Q_{ij}(\omega) = \int_{\rm Antenna}\,x_ix_j\,\varrho({\bf x},\omega)\,d^3x
\end{equation}
and $\Omega_{BD}\/$ is Brans--Dicke's parameter.

Explicit calculation shows that $P_{\rm scalar-tensor}$ is made up of two
contributions:

\begin{equation}
P_{\rm scalar-tensor} = P_{00} + P_{20}
\label{3.1}
\end{equation}
where $P_{00}$ is the scalar, or monopole contribution to the emitted power,
while $P_{20}$ comes from the central quadrupole mode which, as discussed in
\cite{bian-98} and \cite{lobo-95}, is excited together with monopole in JBD
theory. One must however recall that monopole and quadrupole modes of the
sphere happen at {\it different frequencies\/}, so that cross sections for
them only make sense if defined separately. More precisely,

\begin{eqnarray}\label{3.2}
\sigma_{n0}(\omega) & = & \frac{10\pi\,\eta_{n0}\,c^2}{\omega^2}\,
  \frac{\Gamma_{n0}^2/4}{(\omega - \omega_{n0})^2 + \Gamma_{n0}^2/4}
  \label{3.2a} \\
\sigma_{n2}(\omega) & = & \frac{10\pi\,\eta_{n2}\,c^2}{\omega^2}\,
  \frac{\Gamma_{n2}^2/4}{(\omega - \omega_{n2})^2 + \Gamma_{n2}^2/4}
  \label{3.2b}
\end{eqnarray}
where $\eta_{n0}$ and $\eta_{n2}$ are defined like in (\ref{2.5}), with all
terms referring to the corresponding modes. After some algebra one finds that

\begin{eqnarray}\label{3.3}
\sigma_{n0}(\omega) & = & H_n\,\frac{GMv_S^2}{(\Omega_{BD}+2)\,c^3}\,
  \frac{\Gamma_{n0}^2/4}{(\omega - \omega_{n0})^2 + \Gamma_{n0}^2/4}
  \label{3.3a} \\
\sigma_{n2}(\omega) & = & F_n\,\frac{GMv_S^2}{(\Omega_{BD}+2)\,c^3}\,
  \frac{\Gamma_{n2}^2/4}{(\omega - \omega_{n2})^2 + \Gamma_{n2}^2/4}
  \label{3.3b}
\end{eqnarray}

Here, I have defined the dimensionless quantities

\begin{eqnarray}\label{3.4}
  H_n & = & \frac{4\pi^2}{ 9\,(1+\sigma_P)}\,(k_{n0}b_{n0})^2	\label{3.4a} \\
  F_n & = & \frac{8\pi^2}{15\,(1+\sigma_P)}\,(k_{n2}b_{n2})^2	\label{3.4b}
\end{eqnarray}
where $\sigma_P\/$ represents the sphere material's Poisson ratio (most
often very close to a value of 1/3), and the $b_{nl}\/$ are defined in
(\ref{2.2}); $v_S\/$ is the speed of sound in the material of the sphere.

In tables \ref{t.1} and \ref{t.2} I give a few numerical values of the
above cross section coefficients.
\begin{table}
\label{t.1}
\begin{tabular}{|r|r|r|r|r|}\hline
$\ \ \varsigma$ & $n$ & $\ \ k_{n0}^SR$ & \qquad\ \ $D_{n0}$ & \qquad $H_n$ \\
\hline
0.01 & 1 & 5.48738 & -1.43328$\cdot 10^{-4}$ & 0.90929 \\ \hline
     & 1 & 12.2332 & -1.59636$\cdot 10^{-3}$ & 0.14194 \\ \hline
     & 2 & 18.6321 & -5.58961$\cdot 10^{-3}$ & 0.05926 \\ \hline
     & 4 & 24.9693 & -0.001279 		     & 0.03267 \\ \hline
0.10 & 1 & 5.45410 & -0.014218 & 0.89530 \\ \hline
     & 1 & 11.9241 & -0.151377 & 0.15048 \\ \hline
     & 2 & 17.7277 & -0.479543 & 0.04922 \\ \hline
     & 4 & 23.5416 & -0.859885 & 0.04311 \\ \hline
0.25 & 1 & 5.04842 & -0.179999 & 0.73727 \\ \hline
     & 2 & 10.6515 & -0.960417 & 0.30532 \\ \hline
     & 3 & 17.8193 & -0.425087 & 0.04275 \\ \hline
     & 4 & 25.8063 &  0.440100 & 0.06347 \\ \hline
0.50 & 1 & 3.96914 & -0.631169 & 0.49429 \\ \hline
     & 2 & 13.2369 &  0.531684 & 0.58140 \\ \hline
     & 3 & 25.4531 &  0.245321 & 0.01728 \\ \hline
     & 4 & 37.9129 &  0.161117 & 0.07192 \\ \hline
0.75 & 1 & 3.26524 & -0.901244 & 0.43070 \\ \hline
     & 2 & 25.3468 &  0.188845 & 0.66284 \\ \hline
     & 3 & 50.3718 &  0.093173 & 0.00341 \\ \hline
     & 4 & 75.469  &  0.061981 & 0.07480 \\ \hline
0.90 & 1 & 2.98141 & -0.963552 & 0.42043 \\ \hline
     & 2 & 62.9027 &  0.067342 & 0.67689 \\ \hline
     & 3 & 125.699 &  0.033573 & 0.00047 \\ \hline
     & 4 & 188.519 &  0.022334 & 0.07538 \\ \hline
\end{tabular}
\caption{Eigenvalues $k_{n0}^SR\/$, relative weights $D_{n0}$ and $H_n\/$
coefficients for a hollow sphere with Poisson ratio $\sigma_P$\,=\,1/3.
Values are given for a few different thickness parameters $\varsigma$.}
\end{table}

\begin{table}
\label{t.2}
\begin{tabular}{|r|r|r|r|r|r|r|}\hline
$\ \ \varsigma$ & $n$ & $\ \ k_{n2}^SR$ & \qquad\ \ $K_{n2}$ &
 \qquad\ \ $D_{n2}$ &  \qquad\ \ $\tilde D_{n2}$ & \qquad $F_n$ \\
\hline
0.10 & 1 &  2.63836 &  0.855799 & 0.000395 & -0.003142 & 2.94602 \\ \hline
     & 2 &  5.07358 &  0.751837 & 0.002351 & -0.018451 & 1.16934 \\ \hline
     & 3 & 10.96090 &  0.476073 & 0.009821 & -0.071685 & 0.02207 \\ \hline
0.25 & 1 &  2.49122 &  0.606536 &  0.003210 & -0.02494 & 2.55218 \\ \hline
     & 2 &  4.91223 &  0.647204 &  0.019483 & -0.13867 & 1.55022 \\ \hline 
     & 3 &  8.24282 & -1.984426 & -0.126671 &  0.67506 & 0.05325 \\ \hline
     & 4 & 10.97725 &  0.432548 & -0.012194 &  0.02236 & 0.03503 \\ \hline
0.50 & 1 &  1.94340 &  0.300212 &  0.003041 & -0.02268 & 1.61978 \\ \hline 
     & 2 &  5.06453 &  0.745258 &  0.005133 & -0.02889 & 2.29572 \\ \hline 
     & 3 & 10.11189 &  1.795862 & -1.697480 &  2.98276 & 0.19707 \\ \hline
     & 4 & 15.91970 & -1.632550 & -1.965780 & -0.30953 & 0.17108 \\ \hline
0.75 & 1 &  1.44965 &  0.225040 &  0.001376 & -0.01017 & 1.15291 \\ \hline
     & 2 &  5.21599 &  0.910998 & -0.197532 &  0.40944 & 1.82276 \\ \hline
     & 3 & 13.93290 &  0.243382 &  0.748219 & -3.20130 & 1.08952 \\ \hline
     & 4 & 23.76319 &  0.550278 & -0.230203 & -0.81767 & 0.08114 \\ \hline
0.90 & 1 &  1.26565 &  0.213082 &  0.001019 & -0.00755 & 1.03864 \\ \hline 
     & 2 &  4.97703 &  0.939420 & -0.323067 &  0.52279 & 1.54106 \\ \hline
     & 3 & 31.86429 &  6.012680 & -0.259533 &  4.05274 & 1.46486 \\ \hline
     & 4 & 61.29948 &  0.205362 & -0.673148 & -1.04369 & 0.13470\\ \hline 
\end{tabular}
\caption{Eigenvalues $k_{n2}^SR\/$, relative weights $K_{n2}$, $D_{n2}$,
$\tilde D_{n2}$ and $F_n\/$ coefficients for a hollow sphere with Poisson
ratio $\sigma_P$\,=\,1/3. Values are given for a few different thickness
parameters $\varsigma$.}
\end{table}

As already stressed in reference \cite{vega-98}, one of the main advantages
of a hollow sphere is that it enables to reach good sensitivities at lower
frequencies than a solid sphere. For example, a hollow sphere of the same
material and mass as a solid one ($\varsigma$\,=\,0) has eigenfrequencies
which are smaller by

\begin{equation}
 \omega_{nl}(\varsigma) = \omega_{nl}(\varsigma=0)\,(1-\varsigma^3)^{1/3}
 \label{3.5}
\end{equation}
for any mode indices $n\/$ and $l\/$. I now consider the detectability of
JBD GW waves coming from several interesting sources with a hollow sphere.

\subsection{Detectability of JBD signals}

The values of the coefficients $F_n\/$ and $H_n\/$, together with the
expressions (\ref{3.2}) for the cross sections of the hollow sphere, can
be used to estimate the maximum distances at which a coalescing compact
binary system and a gravitational collapse event can be seen with such
detector. I consider these in turn.

\subsubsection{Binary systems}

I consider as a source of GWs a binary system formed by two neutron stars,
each of them with a mass of $m_1$\,=\,$m_2$\,=\,1.4\,$M_\odot$.
The {\it chirp mass\/} corresponding to this system is
$M_c\/$\,$\equiv$\,$(m_1m_2)^{3/5}\,(m_1+m_2)^{-1/5}$\,=\,1.22\,$M_\odot$,
and $\nu_{\rm [5\ cycles]}$\,=\,1270 Hz\footnote{
The frequency $\nu_{\rm [5\ cycles]}$ is the one the system has when it is
5 cycles away from coalescence. It is considered that beyond this frequency
disturbing effects distort the simple picture of a clean binary system
---see \protect\cite{covi-96} for further references.}. Repeating the analysis
carried on in  section five of \cite{brun-99} I find a formula for the minimum distance at 
which a measurement can be performed given a certain signal to noise ratio (SNR), for a {\it quantum limited\/} detector

\begin{eqnarray}\label{4.1}
 r(\omega_{n0}) & = &\left[\frac{5\cdot 2^{1/3}}{32}\,
  \frac{1}{(\Omega_{BD}+2)(12\Omega_{BD}+19)}\,
  \frac{G^{5/3}M_c^{5/3}}{c^3}\,\frac{Mv_S^2}{\hbar\omega_{n0}^{4/3}SNR}\,H_n
  \right]^{1/2}  \label{4.1a}  \\
 r(\omega_{n2}) & = &\left[\frac{5\cdot 2^{1/3}}{192}
  \frac{1}{(\Omega_{BD}+2)(12\Omega_{BD}+19)}\,
  \frac{G^{5/3}M_c^{5/3}}{c^3}\,\frac{Mv_S^2}{\hbar\omega_{n2}^{4/3}SNR}\,F_n
  \right]^{1/2}  \label{4.1b}
\end{eqnarray}

For a CuAL sphere, the speed of sound is $v_S\/$\,=\,4700 m/sec. I report
in table \ref{t.3} the maximum distances at which a JBD binary can be seen
with a 100 ton hollow spherical detector, including the size of the sphere
(diameter and thickness factor) for $SNR=1$. The Brans-Dicke parameter $\Omega_{BD}\/$
has been given a value of 600. This high value has as a consequence that
only relatively nearby binaries can be scrutinised by means of their scalar
radiation of GWs. A slight improvement in sensitivity is appreciated as the
diameter increases in a fixed mass detector. Vacancies in the tables mean
the corresponding frequencies are higher than $\nu_{\rm [5\ cycles]}$.

\begin{table}

\label{t.3}
\begin{tabular}{|r|r|r|r|r|r|}\hline
$\varsigma $ & $\Phi$ (m) & $\nu_{10}$(Hz) & $\nu_{12}$ (Hz)
& $r(\nu_{10})$ (kpc) & $r(\nu_{12})$ (kpc) \\ \hline
0.00 & 2.94 & 1655 & 807 & $-$ & 29.8 \\ \hline
0.25 & 2.96 & 1562 & 771 & $-$ & 30.3 \\ \hline
0.50 & 3.08 & 1180 & 578 & 55  & 31.1 \\ \hline
0.75 & 3.5  & 845  & 375 & 64  & 33   \\ \hline
0.90 & 4.5  & 600  & 254 & 80  & 40\\ \hline
\end{tabular}
\caption{Eigenfrequencies, sizes and distances at which coalescing binaries
can be seen by monitoring of their emitted JBD GWs. Figures correspond to a
100 ton CuAl hollow sphere.}
\end{table}

\begin{table}
\label{t.4}
\begin{tabular}{|r|r|r|r|r|r|}\hline
$\varsigma $ & $M$ (ton) & $\nu _{10}$(Hz) & $\nu_{12}$(Hz) &
$r(\nu_{10})$ (kpc) & $r(\nu_{12})$ (kpc) \\ \hline
0.00 & 105   & 1653 & 804 & $-$  & 33   \\ \hline 
0.25 & 103.4 & 1541 & 760 & $-$  & 31   \\ \hline
0.50 & 92    & 1212 & 593 & 52   & 27.6 \\ \hline
0.75 & 60.7  & 997  & 442 & 44.8 & 23   \\ \hline
0.90 & 28.4  & 910  & 386 & 32   & 16.3\\  \hline
\end{tabular}
\caption{Eigenfrequencies, sizes and distances at which coalescing binaries
can be seen by monitoring of their emitted JBD GWs. Figures correspond to a
3 metres external diameter CuAl hollow sphere.}
\end{table}

\subsubsection{Gravitational collapse}

The signal associated to a gravitational collapse has recently been modeled,
within JBD theory, as a short pulse of amplitude $b\/$, whose value can be
estimated as \cite{shibata}:

\begin{equation}
b \simeq 10^{-23}\,\left(\frac{500}{\Omega_{BD}}\right)
  \left(\frac{M_*}{M_\odot}\right)\left(\frac{10\,Mpc}{r}\right)
\label{4.2}
\end{equation}
where $M_*$ is the collapsing mass.

The minimum value of the Fourier transform of the amplitude of the scalar
wave, for a quantum limited detector at unit signal-to-noise ratio, is
given by \cite{bian-98}

\begin{equation}
\left|b(\omega_{nl})\right|_{\min } = \left(
\frac{4\hbar}{Mv_S^2\omega_{nl}K_n}\right)^{1/2}
\label{4.3}
\end{equation}
where $K_n=2H_n$ for the mode with $l=0$ and $K_n=F_n/3$ for the 
mode with $l=2, m=0$.

The duration of the impulse, $\tau \approx 1/f_c$, is much shorter than
the decay time of the $nl\/$ mode, so that the relationship between $b\/$
and $b(\omega_{nl})$ is

\begin{equation}
b \approx | b(\omega_{nl})| f_c\ \ at \\
  \omega_{nl} = 2\pi f_c
  \label{4.4}
\end{equation}
so that the minimum scalar wave amplitude detectable is

\begin{equation}
 |b|_{\min}\approx\left(
  \frac{4\hbar\omega_{nl}}{\pi^2Mv_S^2K_n}\right)^{1/2}
  \label{4.5}
\end{equation}

Let us now consider a hollow sphere made of molibdenum, for which the speed
of sound is as high as $v_S\/$\,=\,5600 m/sec. For a given detector mass and
diameter, equation (\ref{4.5}) tells us which is the minimum signal detectable
with such detector. For example, a solid sphere of $M=31$ tons and 1.8 metres in
diameter, is sensitive down to $b_{\min}$\,=\,1.5\,$\cdot$\,10$^{-22}$.
Equation (\ref{4.2}) can then be inverted to find which is the maximum
distance at which the source can be identified by the scalar waves it emits.
Taking a reasonable value of $\Omega_{BD}\/$\,=\,600, one finds that
$r(\nu_{10})$\,$\approx$\,0.6 Mpc.

Like before, I report in tables \ref{t.4}, \ref{t.5} and \ref{t.6} the 
sensitivities
of the detector and consequent maximum distance at which the source appears
visible to the device for various values of the thickness parameter
$\varsigma\/$. In table~\ref{t.5} a detector of mass of 31 tons has
been assumed for all thicknesses, and in tables~\ref{t.4}, \ref{t.6} a 
constant outer
diameter of 3 and 1.8 metres has been assumed in all cases.

\begin{table}
\label{t.5}
\begin{tabular}{|r|r|r|r|r|}\hline
$\varsigma $ & $\phi$ (m) & $\nu_{10}$ (Hz) & $|b|_{\min }$ (10$^{-22}$)
& $r(\nu_{10})$ (Mpc) \\ \hline
0.00 & 1.80 & 3338 & 1.5  & 0.6  \\ \hline
0.25 & 1.82 & 3027 & 1.65 & 0.5  \\ \hline
0.50 & 1.88 & 2304 & 1.79 & 0.46 \\ \hline
0.75 & 2.16 & 1650 & 1.63 & 0.51 \\  \hline
0.90 & 2.78 & 1170 & 1.39 & 0.6\\ \hline
\end{tabular}
\caption{Eigenfrequencies, maximum sensitivities and distances at which
a gravitational collapse can be seen by monitoring the scalar GWs it emits.
Figures correspond to a 31 ton Mb hollow sphere.}
\end{table}

\begin{table}
\label{t.6}
\begin{tabular}{|r|r|r|r|r|}\hline
$\varsigma $ & $M$ (ton) & $\nu_{10}$ (Hz) & $|b|_{\min }$ (10$^{-22}$)
& $r(\nu_{10})$ (Mpc) \\ \hline
0.00 & 31.0  & 3338 & 1.5  & 0.6  \\ \hline 
0.25 & 30.52 & 3062 & 1.71 & 0.48 \\ \hline  
0.50 & 27.12 & 2407 & 1.95 & 0.42 \\ \hline  
0.75 & 17.92 & 1980 & 2.34 & 0.36 \\ \hline  
0.90 & 8.4   & 1808 & 3.31 & 0.24 \\   \hline
\end{tabular}
\caption{Eigenfrequencies, maximum sensitivities and distances at which
a gravitational collapse can be seen by monitoring the scalar GWs it emits.
Figures correspond to a 1.8 metres outer diameter Mb hollow sphere.}
\end{table}

\section{Deviations from gravity at short distances}
There has been much interest recently for compatifications of string 
theory models which could lead to effects of quantum gravity at the scale
of the TeV. This idea is not new, but has received new impulse from the
works of Ref.\cite{add, aadd}. Let me recall very briefly the 
main idea.
The low-energy action of the heterotic string compactified to four 
dimensions looks like
\begin{equation}
S=\int d^4x {V\over \lambda^2_H}(l_H^{-8}R+l_H^{-6}F^2)
\label{a1}
\end{equation}
where $\lambda_H=g\sqrt{V}/l^3_H$ is the dimensionless string coupling
(the exponential of the vacuum expectation value of the dilaton),
$l_H$ is the inverse of the mass scale, the subscript $H$ refers to the fact
I am considering a 
heterotic string, $V$ is the compactified volume, $g$ is the gauge coupling 
constant and, for the sake of simplicity, in (\ref{a1}) I have only considered
the gauge and gravitational sectors. From these expressions I see that for
typical values of $g\approx 1/5$, $M_H$ is of the order of the
Planck scale
and the theory is weakly coupled for $V\approx \lambda_H^6$.
Let me now esamine the situation for the case of strings of type I.
This is a theory of open and closed strings. The closed string sector
generates gravity while the gauge sector is generated by open strings
whose end are confined to propagate on D-branes.
Having in mind a four dimensional compactification I divide the six
internal (compactified) dimensions as parallel or transvers to the 
D-brane. Assuming that the standard model is localized on a $p\ge 3$
brane, there are $p-3$ longitudinal and $9-p$ transverse
directions.
The corresponding low-energy effective action for the zero-mass
sector looks like
\begin{equation}
S=\int d^{10}x {1\over \lambda^2_I l_I^8}R+
\int d^{p+1}x {l_I^{3-p}\over \lambda_I}F^2.
\label{a2}
\end{equation}
Upon compactification, the Planck lenght and gauge coupling
are given by
\begin{equation}
{1\over \lambda^2_I}={V_\parallel V_\perp\over \lambda^2_I l_I^8}, \ \ 
{1\over g^2}={l_I^{3-p}V_{\parallel}\over \lambda_I} 
\label{a3}
\end{equation}
where $V_\parallel (V_\perp)$ is the compactified volume parallel (transverse) 
to the p-brane. The requirement of weak coupling ($\lambda_I< 1$)
implies $V_\parallel\approx l^{p-3}$ while the transverse volume is unrestricted.
Then the relation between the Planck scale and that of the string of type I
becomes
\begin{equation}
M^2_P={M_I^{n+2}R_\perp^n\over g^4 v_\parallel }, \ \ 
\lambda_I= g^2 v_\parallel.
\label{a4}
\end{equation}
Here $v_\parallel > 1$ is the parallel volume in string units and
I have assumed that the transverse compactified space of dimension
$n=9-p$ is isotropic. From (\ref{a4}) I see that choosing $n$ and 
$R_\perp$ in a suitable way it is possible to find a $M_I\ll M_P$.
Furthermore (\ref{a4}) can be seen as coming from the Gauss's law
for gravity in $4+n$ dimensions thus leading to a Newton's constant
\begin{equation}
G_N^{(4+n)}=g^4 l_I^{n+2}v_\parallel .
\end{equation}
For $M_I\approx 1$ TeV, I find $R_\perp=10^8$ km, $.1$ mm ($10^{-3}$ eV), 
$.1$ fermi ($10$ MeV)
for $n=1, 2, 6$ large dimensions. $n=2$ is still not ruled out from
experimental data. It is thus considered important to perform experiments
to check the consistency of gravity at the mm scale, below which the effect
of extra-dimension should be considered important \cite{add}.

As the prototype experiment\footnote{See Ref.\cite{fisch} for a detailed
description of experimental measures of the so-called fifth force.}
to perform such a measurement I take the setting
of Fig.1, in which a test mass (generator) is made to oscillate in front of
another oscillator (detector). The frequency of oscillation of the generator
is the same as the frequency of the first resonant mode of the detector.
This frequency is typically of the order of the KHz to try to decouple the system
from background noise. It is also advisable to keep the entire apparatus
at very low temperature to decrease the noise.
%%%%%%%%%%%%%%%%%%%%%%%%%%%%%%%%%%%%%%%%%%%%%%%%%%%%%%%%%%%%%%%%%
\begin{figure}[htb]
\centerline{\hbox{ \psfig{figure=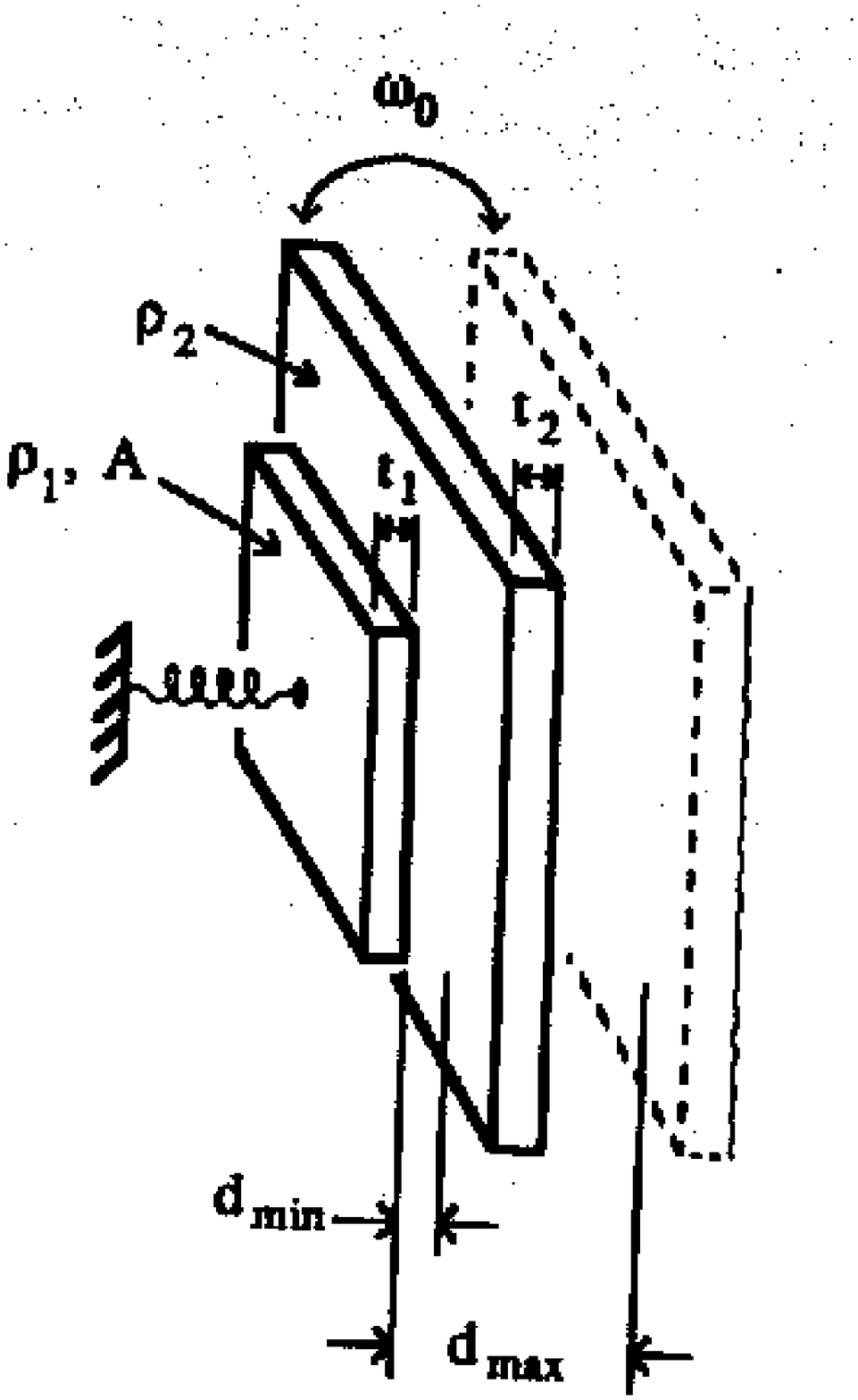} }}
\caption{\footnotesize Configuration of the prototype experiment discussed 
in the text.  \label{fig}}
\end{figure}
%%%%%%%%%%%%%%%%%%%%%%%%%%%%%%%%%%%%%%%%%%%%%%%%%%%%%%%%%%%%%%%%%
The main advantage of a configuration of the type of that of Fig.1
is that it is a null gravity experiment: {\it i.e.} gravity
\footnote{I mean the part of the potential which goes like the
inverse of the distance between two test masses.} is constant
between the two parallel plates. In the Fourier transformed
domain I am thus lead to a signal which is different from zero only for 
deviations
from gravity which, as usual, are parametrized by a Yukawa type potential
\begin{equation}
V(r)=-G_N^{(4)}\alpha\int_{V_d} d^3x {e^{-{r\over\lambda}}\over r}
\end{equation}
which expresses the interaction of an atom of the generator plate (of area 
$A$, thickness $t_1$, density $\rho_1$) 
with the detector plate which is taken to have geometrical dimensions such
that its area is much bigger than $A$.
Integrating over the generator plate and taking the derivative with respect
to the distance between the two parallel plates, I find the force 
\begin{equation}
F_Y=-2\pi\alpha G_N^{(4)}\rho_1\rho_2\lambda^2 Ae^{-{d(t)\over\lambda}}
(1-e^{-{t_1\over\lambda}})(1-e^{-{t_2\over\lambda}}),
\label{a5}
\end{equation}
where $d(t)=\bar d+d_0\sin\omega_0 t$ is the distance between the two plates
with $d_0=(d_{max}-d_{min})/2$ and $\bar d=(d_{max}+d_{min})/2$.
Fourier transforming (\ref{a5}) leads to
\begin{equation}
\widetilde F_Y=-2\pi\alpha G_N^{(4)}\rho_1\rho_2 A\lambda^2 I_1({d_0\over\lambda})
e^{-{\bar d\over\lambda}}(1-e^{-{t_1\over\lambda}})(1-e^{-{t_2\over\lambda}}),
\label{a6}
\end{equation}
where $I_1$ is a Bessel function. To obtain (\ref{a6}) I have used the fact that
only the oscillator first resonant mode is relevant for our discussion.
An estimate of the noise is given by
\begin{equation}
F_{noise}=\sqrt{kTm\omega_0\over Q\tau},
\label{a7}
\end{equation}
where m is the mass of the detector, k is the Boltzmann constant, 
T the temperature
at which the experiment is performed, Q is the detector's quality factor 
and $\tau$ the integration time. Plugging into (\ref{a7}) typical values 
for the quantities involved, I get $F_{noise}\approx 10^{-10}$ dyne. 
The ratio between  (\ref{a6}) and (\ref{a7})
gives the signal to noise ratio, SNR. A proposed experiment along these lines
has been described in \cite{lcp}.    

Let me now examine what are the main sources of noise that have to be
monitored in such an experiment:
\begin{itemize}
\item{} Casimir type forces
\item{} Surface type forces
\end{itemize}

Fourier transforming the Casimir force between two parallel plates,
I get
\begin{equation}
\widetilde F_{Casimir}=-{\pi^2 \hbar c A\over 280}{(d_0^3+4d_0^2 \bar d)\over
(d_0^2+\bar d^2)^{7/2}}
\label{a8}
\end{equation}
For values typical of these experiments (probing gravity at the mm scale)
$\widetilde F_{Casimir}\approx 10^{-12}$ dyne that is a value which is a couple
of order of magnitude smaller than our noise. Consequently I  do not have to worry
about corrections to (\ref{a8}) coming from lack of parallelism between the plates,
the temperature being different from zero, the materials being not perfect metals
or the two plates being not at rest.

Let me come now to the second source of noise which I have denoted surface type 
forces because they are generated by the superficial properties of the two plates.
It is in fact well known that if I put in front of each other two materials I
can measure in the space between the two an electric field of the order of the tens
of millivolt.  This field is generated by patches of charges on the surface 
of the materials which, in first istance I attribute to impurities on the 
parallel two surfaces. An estimate of the force generated is given by
\begin{equation}
F_{electric}={1\over 4\pi}{(\Delta V)^2 A\over (\bar d+d_0\sin\omega_0 t)^2}
\approx 10^{-10}{A\over (\bar d+d_0\sin\omega_0 t)^2} {\rm dyne}.
\end{equation}
The strength of $F_{electric}$ is such that it could overcome the signal 
from the Yukawa potential. From the experimental point of view this force is
dealt with by connecting the two material to earth and among them or giving
a bias current to balance for the phenomena. 
Giving the importance  of the consequences of the presence of $F_{electric}$ 
it is then time to better understand the physics of the problem. If the material
of which the two plates are made are different metals, then the electric field
originates from the difference of the Fermi levels between the two. But even if
the materials are the same the effect is still there, due to the fact that the metals
are not monocrystals. It is anyway possible two coat the two plates with a thin layer
of monocrystal material. Few angstr\"om are sufficient to change the 
superficial properties. In my opinion this is the most elegant way to cope with 
this problem.

In conclusion, in this chapter I have shown that performing experiments of gravity
at the mm scale is possible given our understanding of the possible noise sources.
I hope that very soon we will have experimental data to discuss.

\end{document}